\documentclass[twocolumn]{jpsj3} %% two-column layout
\title{Edge States of Monolayer and Bilayer Graphene Nanoribbons}

\author{Wei {\sc Li} and Ruibao {\sc Tao}}
\inst{Department of Physics, Fudan University, Shanghai 200433, China}

\abst{On the basis of tight-binding lattice model, the
edge states of monolayer and bilayer graphene nanoribbons (GNRs)
with different edge terminations are studied. The effects of
edge-hopping modulation, spin-orbital coupling (SOC), and bias
voltage on bilayer GNRs are discussed. We observe the following: (i) Some new extra edge states can be created by
edge-hopping modulation for monolayer GNRs. (ii) Intralayer
Rashba SOC plays a role in depressing the band energy gap $E_g$
opened by intrinsic SOC for both monolayer and bilayer GNRs. An
almost linear dependent relation, {\it i.e.}, $E_g\sim \lambda_R$, is found.
(iii) Although the bias voltage favors a bulk energy gap for
bilayer graphene without intrinsic SOC, it tends to reduce the
gap induced by intrinsic SOC. (iv) The topological phase of
the quantum spin Hall effect can be destroyed completely by
interlayer Rashba SOC for bilayer GNRs.}

\kword{Edge State, edge-hopping modulation, Spin-Orbit Coupling, Graphene Nanoribbon}

\begin{document}
\maketitle
%**********************************************************************
\section{Introduction\label{sec:intro}}
%**********************************************************************
The electronic properties of graphene\cite{Novoselov1,Novoselov2} have attracted great research interest
in recent years owing to the unconventional physical properties and
remarkable potential in advanced nanoelectronics applications of this material. Differently from
conventional parabolic excitations in semiconductors and metals, the low-energy
excitation of graphene has a linear dispersion relation, and graphene
quasiparticles obey the massless relativistic Dirac-like equation (up to the
order of energy of 1000 K)\cite{Wallace,Peres2} with a Fermi velocity $%
v_{F}\approx 10^{6}$m/s. In addition to various extraordinary physical
behaviors such as the half-integer quantum Hall effect (QHE)\cite%
{YZhang,YBZhangPRL} and Klein tunnelling\cite{Katsnelson}, graphene-based models
studies also triggered active research in the field of a whole new class of topological insulators.

A theoretical model of topological insulators in graphene with exhibiting spin-orbit
coupling (SOC) was previously suggested by Kane and Mele\cite{Kane1,
Kane2}, where two kinds of SOC were introduced: (i) Intrinsic SOC,
induced by the next-nearest-neighbor (n.n.n.) spin-dependent
hopping, which is similar to Haldane's toy model of QHE without
an external magnetic field\cite{Haldane}; (ii) Rashba
SOC\cite{Hongki, zhenhuaQiao, Mahdi2, Mahdi}, induced by the
structure inversion asymmetry and described by the spin-dependent
nearest-neighbor (n.n) hopping. Intrinsic SOC can generate a bulk energy
gap\cite{Ralph}. As a result, graphene can be driven into a topological
insulator phase [quantum spin Hall effect (QSHE) in this case] from a two-dimensional (2D) gapless
semiconductor phase. Such a topological phase is robust to nonmagnetic
disorder, while it can be continually suppressed by an increase in the
strength of Rashba SOC. However, according to the
later work by Yao {\it et al}.\cite{YuguiYao}, SOC (on the order of strength of $%
10^{-3}$meV) in graphene is much weaker than the one suggested in
Kane and Mele's model\cite{Kane1}. Therefore, SOC in graphene has been
neglected in many works. However, it is unknown if someone
can apply some external sources, such as laser beam radiation and
proximity effect to realize a steady state with a strong SOC
for graphene\cite{Yu2008,Castro2009}. This should be an interesting problem for future
applications. Owning to
the correspondence between bulk and edge states in topological
insulators \cite{YHatsugai}, the study of edge states is
important for understanding the topological properties of such
systems. Thus, investigating the edge states of monolayer graphene
(MG) and bilayer graphene (BG) naturally becomes an interesting
issue. Not only is the study of edge states important for
exploring some fundamental physics, but it is also useful for
future applications, especially if we can find ways to create and
manipulate some new edge states. In this paper, we will focus on
the study of the edge states of MG nanoribbons (MGNRs) and BG
nanoribbons (BGNRs) in detail.

It is helpful to provide a briefly review of what is basically known in these topological insulator-based models. As is well known, a
nondispersive zero edge mode (namely, a flat band with zero energy) appears in MGNRs with a zigzag edge.
If n.n.n hopping is taken into account, such a zero
edge mode is changed into a dispersive one\cite{KSasaki}. On the other hand, there is no edge state in MGNRs with an armchair edge.
The spectrum of low-energy excitation becomes quadratic\cite{Eduardo} in BG. When a bias voltage is applied between two
layers of BG, the gapless spectrum of BG can be opened and tuned by the bias
voltage. Such a transition from a gapless semiconductor to a gapful semiconductor was
observed in a previous experiment\cite{CASTRO}. It was also shown that edge states
exist in BGNRs with a zigzag edge\cite{CastroPRL}, but not
in BGNRs with an armchair edge.

In the present paper, we study the electronic properties of MGNRs and BGNRs.
The roles of SOC and bias voltage in BGNRs, as well as edge-hopping modulation are investigated.
Firstly, we present the results of the edge states of MGNRs with edge-hopping modulation, but no SOC is included. Our calculations show that some new extra edge states can be created and that the energy dispersion of edge states can be modified by edge-hopping modulation. This suggests a way to create and manipulate some new optical
transition and transport channels. Secondly, it is found that both
intralayer Rashba SOC and bias voltage in BGNRs depress the
bulk energy gap opened by intrinsic SOC. Differently
from intralayer Rashba SOC, interlayer Rashba SOC is favorable for
opening a gap and driving a topological \textquotedblleft edge
semimetal\textquotedblright\ to a band insulator.

The rest of this paper is organized as follows. In $\S$\ref{SLGNR_NSOC}, we study the edge states of MGNRs with
edge-hopping modulation, while no SOC is considered. The effect of SOC on
the energy spectrum, particularly on the edge states, of MGNRs is investigated in $\S$\ref{SLGNR_SOC}. In $\S$
\ref{BLGNR_SOC}, we present our results on the edge states of BGNRs. Finally, we give a brief conclusion in $\S$
\ref{conclusions}.

%%%%%%%%%%%%%%%%%%%%%%%%%%%%%%%%%%%%%%%%%%%%%%%%%%%%%%%%%%%%%%%%%%%%%%%%
\section{Edge States in MGNRs without SOC}
\label{SLGNR_NSOC}
%%%%%%%%%%%%%%%%%%%%%%%%%%%%%%%%%%%%%%%%%%%%%%%%%%%%%%%%%%%%%%%%%%%%%%%%
Electronic transport in graphene is mainly contributed by $p_{z}$-$\pi$
orbitals. The theoretical model Hamiltonian based on tight-binding
approximation (TBA) is

\begin{figure}[tbp]
\centering
\includegraphics[width=7.5cm, height=5.5cm]{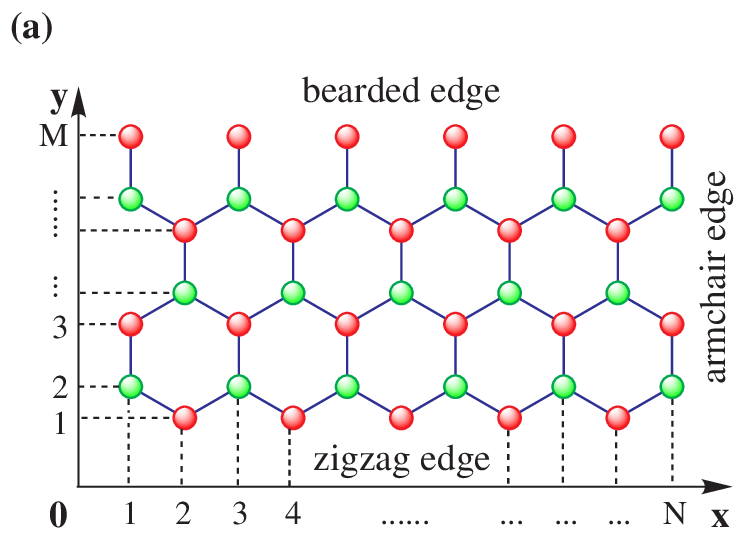}
\hspace{1mm}
\includegraphics[width=7.5cm,
height=5.5cm]{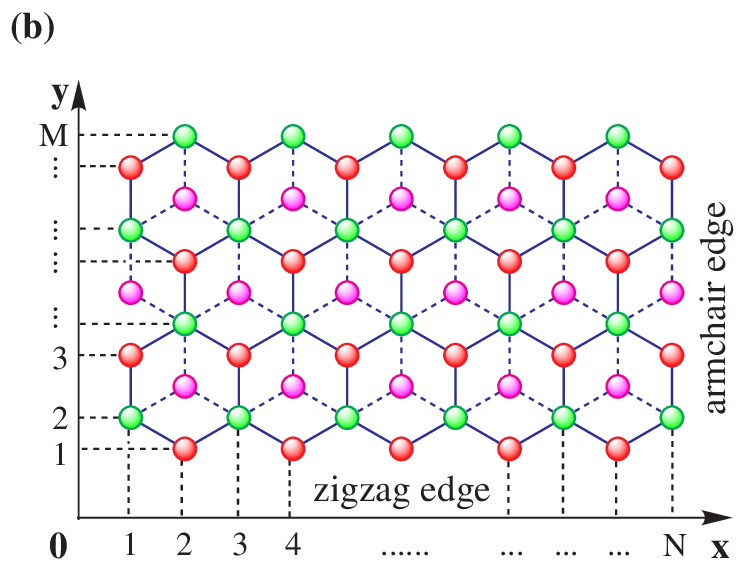}
\caption{(Color online) Schematic
illustration of the lattice structures of MGNRs (a) and BGNRs (b)
with zigzag/bearded and armchair edges, consisting of sublattices A
and B. The widths of each nanoribbon are M and N for the zigzag and
armchair edges, respectively.} \label{fig:one}
\end{figure}

\begin{eqnarray}
{\hat{\mathcal{H}}}_{1} &=&H_{bulk}+H_{edge}\label{eq:one}
\end{eqnarray}
\begin{eqnarray*}
H_{bulk} &=&\sum_{\langle i,j\rangle}tc_{i}^{\dag }c_{j}+\sum_{\langle\langle i,j\rangle\rangle }t^{\prime
}c_{i}^{\dag }c_{j}+h.c.  \\
H_{edge} &=&\sum_{i\in edge}t_{0}c_{i}^{\dag }c_{j}+h.c.
\end{eqnarray*}
where the summation indices of $\langle i,j\rangle$ in $H_{bulk}$ refer to n.n. hopping
terms, while the indices $\langle\langle i,j\rangle\rangle$ refer to n.n.n. pair hopping
terms in the bulk. The hopping constants in bulk are $t\approx 2.8$eV (as the energy unit) and $%
t^{\prime}\approx 0.1$. There have been several reports on edge-hopping modulation
induced at atoms near edge surfaces, such as the edge perturbations\cite%
{Jaroslaw} and local potential applied at the edge
atoms\cite{Niu}. Here, we supply a model of modulation for
the edge hopping constant $t_{0}$. When the edge hopping constant $t_{0}$
does not equal the bulk value $t$, it corresponds to boundary
softening $(t_{0}<t)$ or stiffening $(t_{0}>t)$. Edge-hopping
modulation can be induced by an external field, structure deformation
and element substitution that will change the wave functions at
the edge sites, thus affecting the overlap integrals and modifying the
hopping constant. For example, if one substitutes an
edge atom with a larger valence iron or applies a positive voltage
on the edge atom, conductance electrons will distribute more
closely to iron. As a result, the degree of overlap integration of hopping
decreases, and a softening case is realized. In the stiffening
case, one should substitute an edge atom with a smaller valence
iron or apply a negative voltage on the edge atom. Phenomenologically, we adopt the $n.n.$ hopping
Hamiltonian $H_{edge}$ near edges to simulate some effective
edge-hopping modulation. The indices of $\langle i,j\rangle$ in
$H_{edge}$ refer to n.n. pairs of edge sites. We coordinate
the lattice sites with double indices $(i_{x}, i_{y})$.

\begin{figure}[tbp]
\centering
\includegraphics[bb=1 1 540 430,width=4.1cm,height=3.4cm]{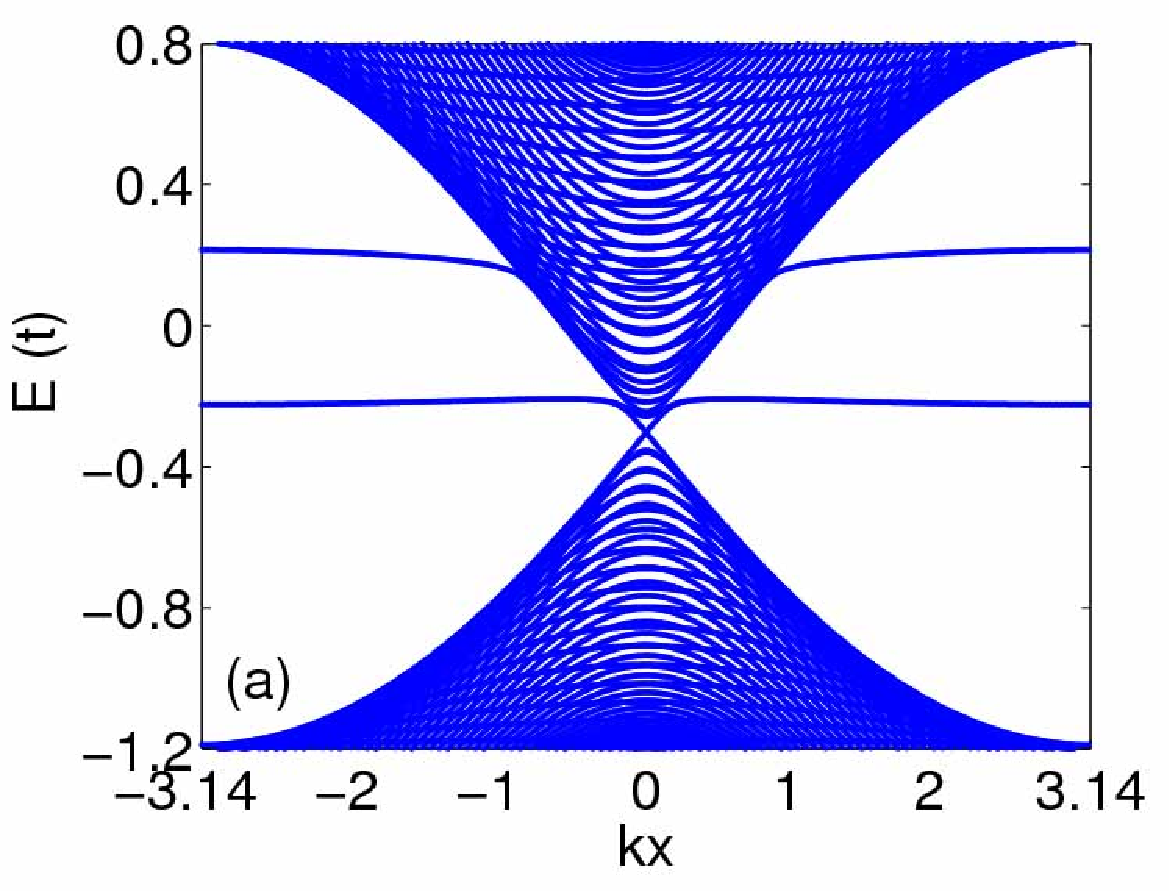} \hspace{1mm} %
\includegraphics[bb=1 1 540 430,width=4.1cm,height=3.4cm]{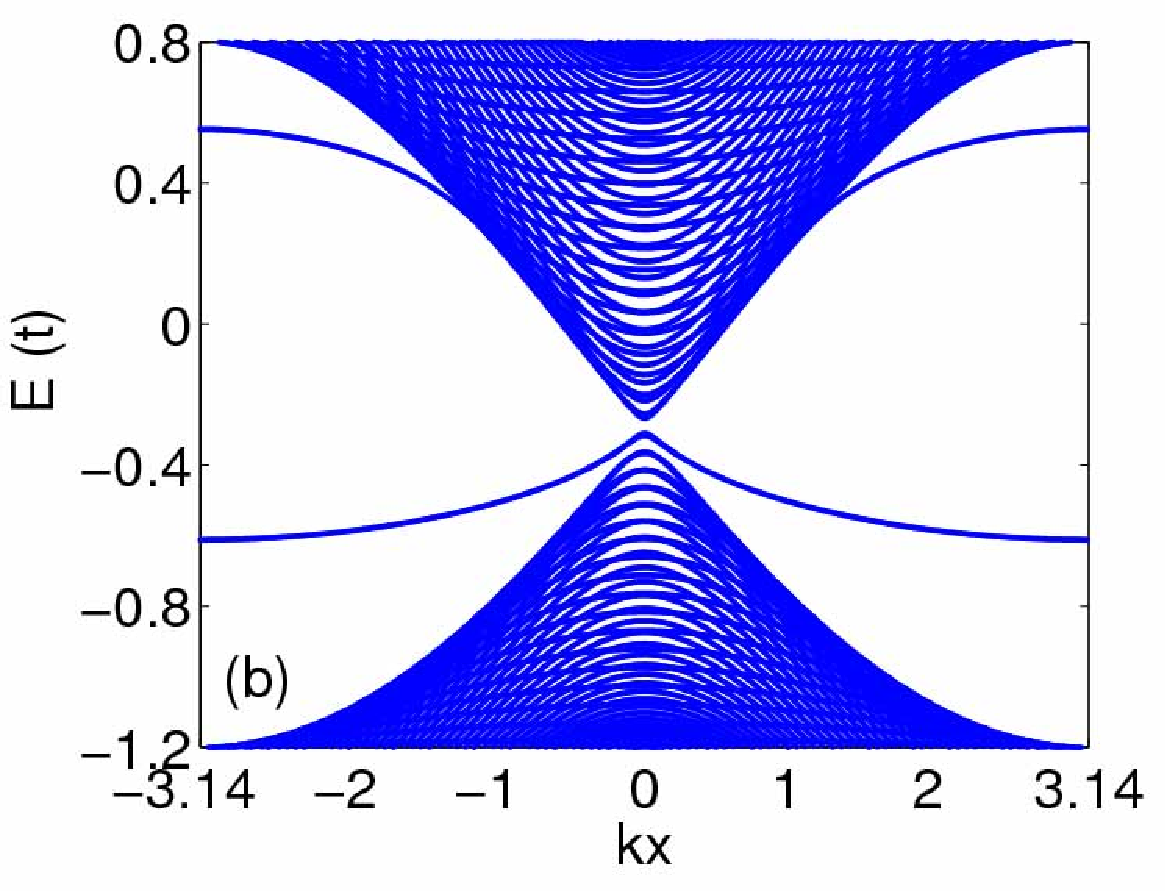}
\hspace{1mm}
\includegraphics[bb=1 1 540 430,width=4.1cm,height=3.4cm]{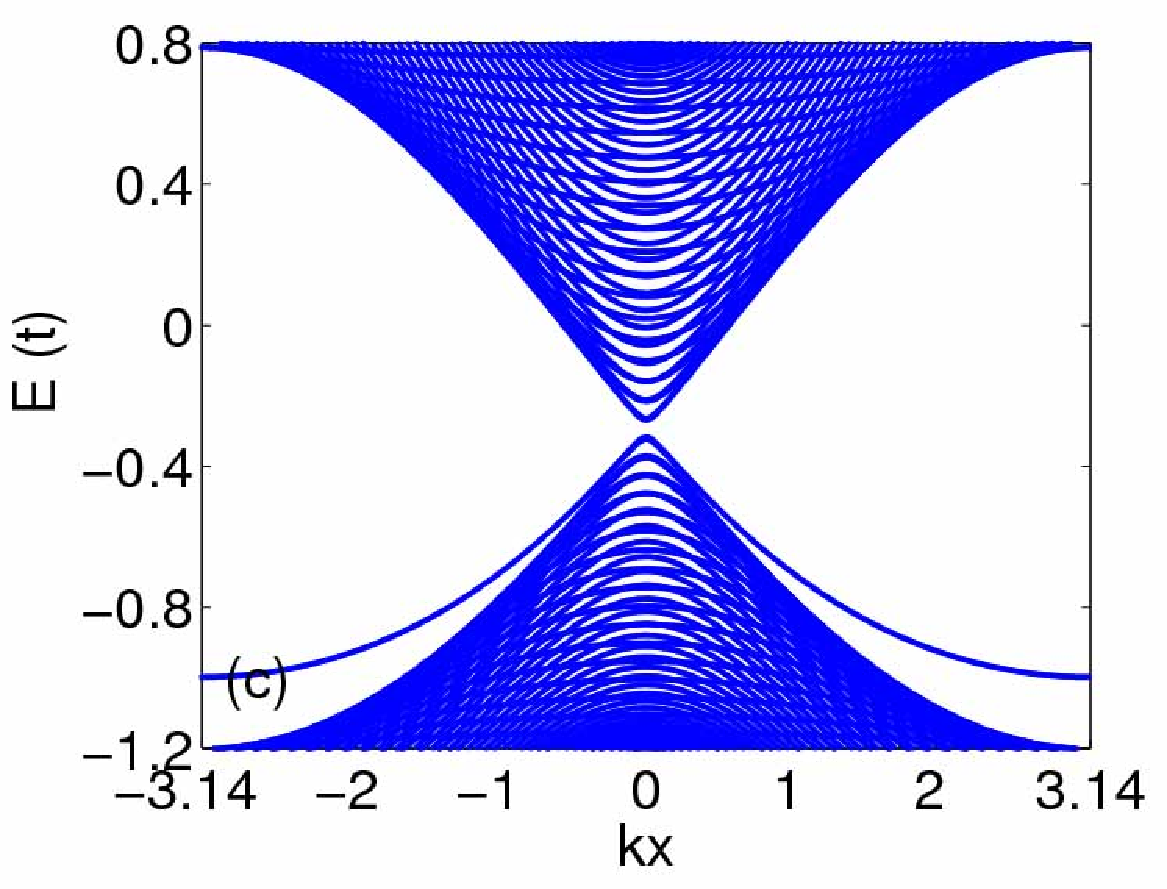} \hspace{1mm} %
\includegraphics[bb=9 1 730 534,width=4.15cm,height=3.3cm]{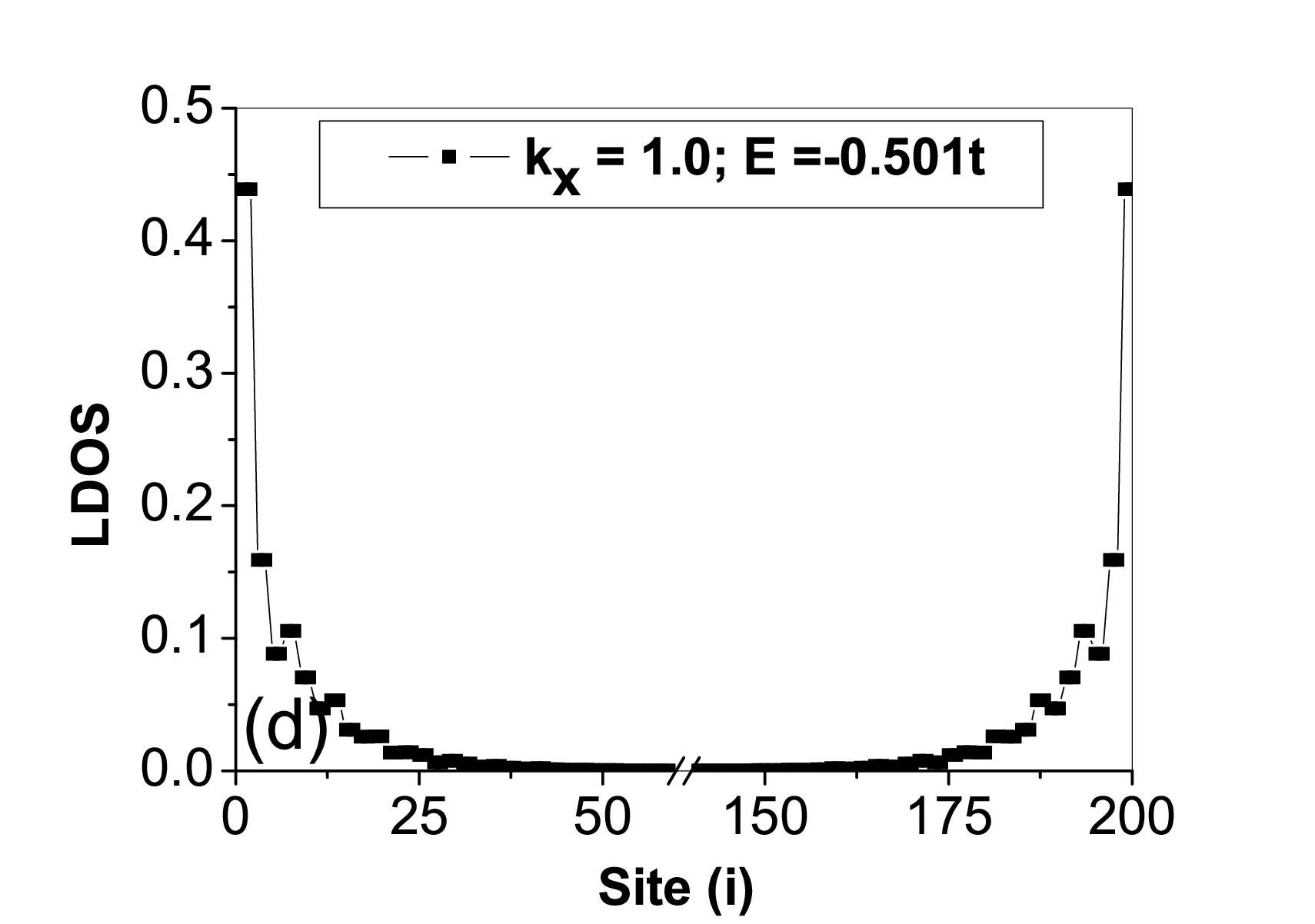}
\caption{(Color online) (a)-(c) Energy band structure of
armchair edged MGNRs, and (d) probability distribution $|\protect\psi|^{2}$ as a
function of position across nanoribbon corresponding to (b).
The parameters we used are as follows: for (a), $t_{0}=0.2$; for (b), $t_{0}=0.5$; and
for (c), $t_{0}=0.8$.} \label{fig:two}
\end{figure}

\begin{figure}[tbp]
\centering
\includegraphics[bb=1 1 540 430,width=4.1cm,height=3.4cm]{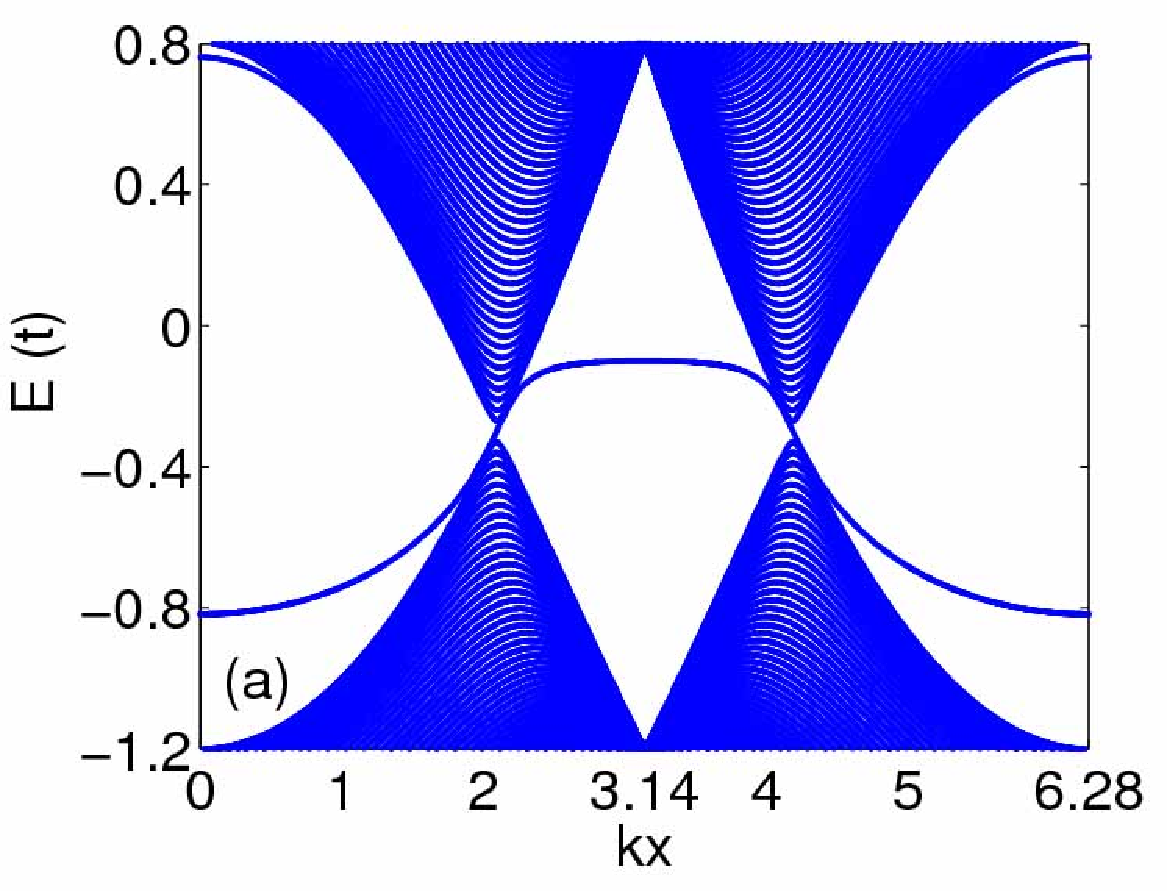} \hspace{1mm}
\includegraphics[bb=1 1 540 430,width=4.1cm,height=3.4cm]{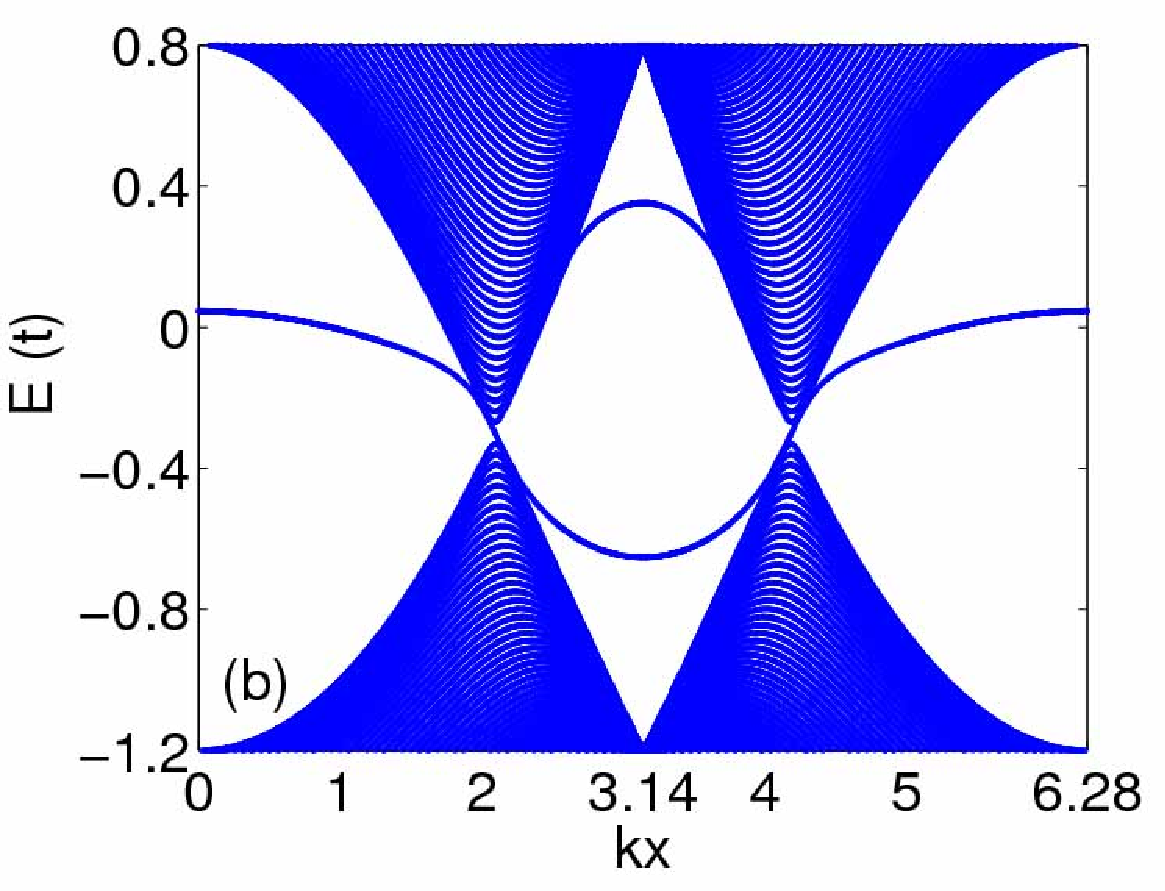} \hspace{1mm} %
\includegraphics[bb=1 1 540 430,width=4.1cm,height=3.4cm]{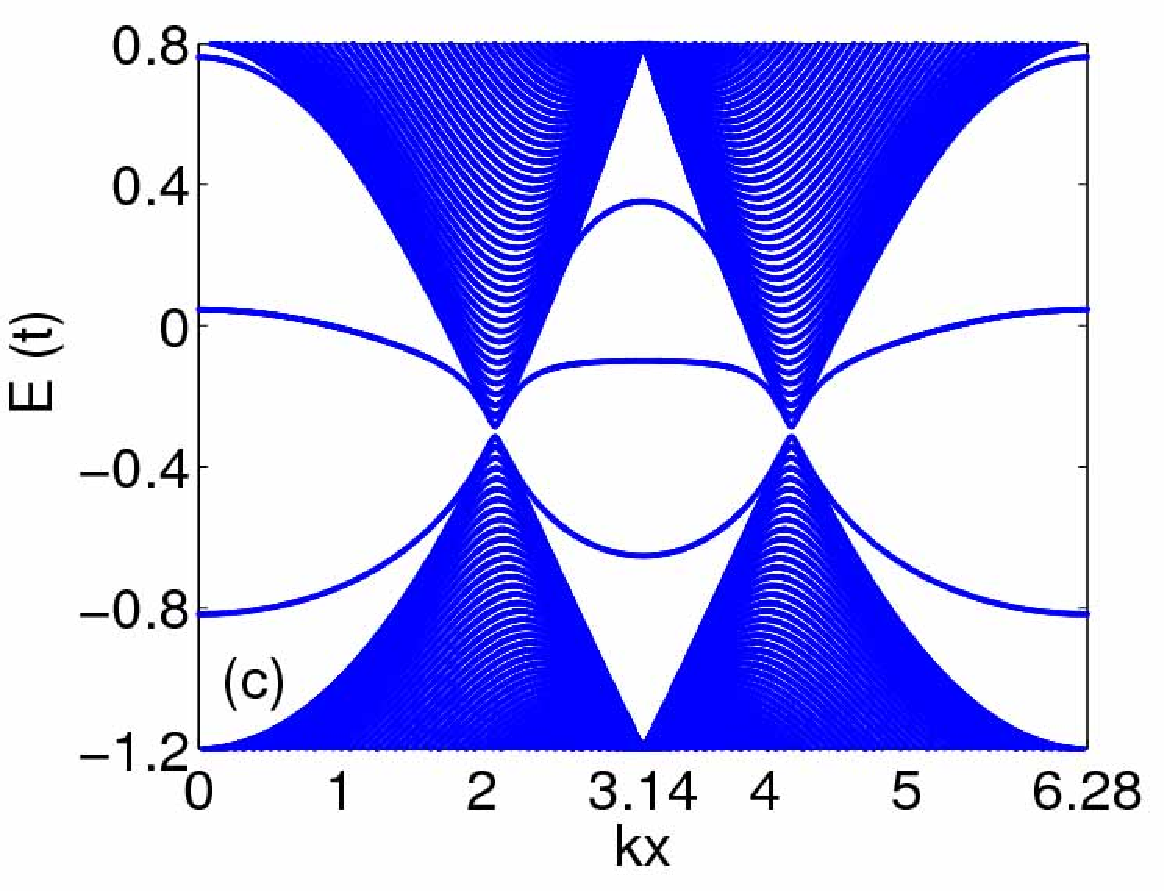}
\hspace{1mm}
\includegraphics[bb=9 1 730 534,width=4.1cm,height=3.35cm]{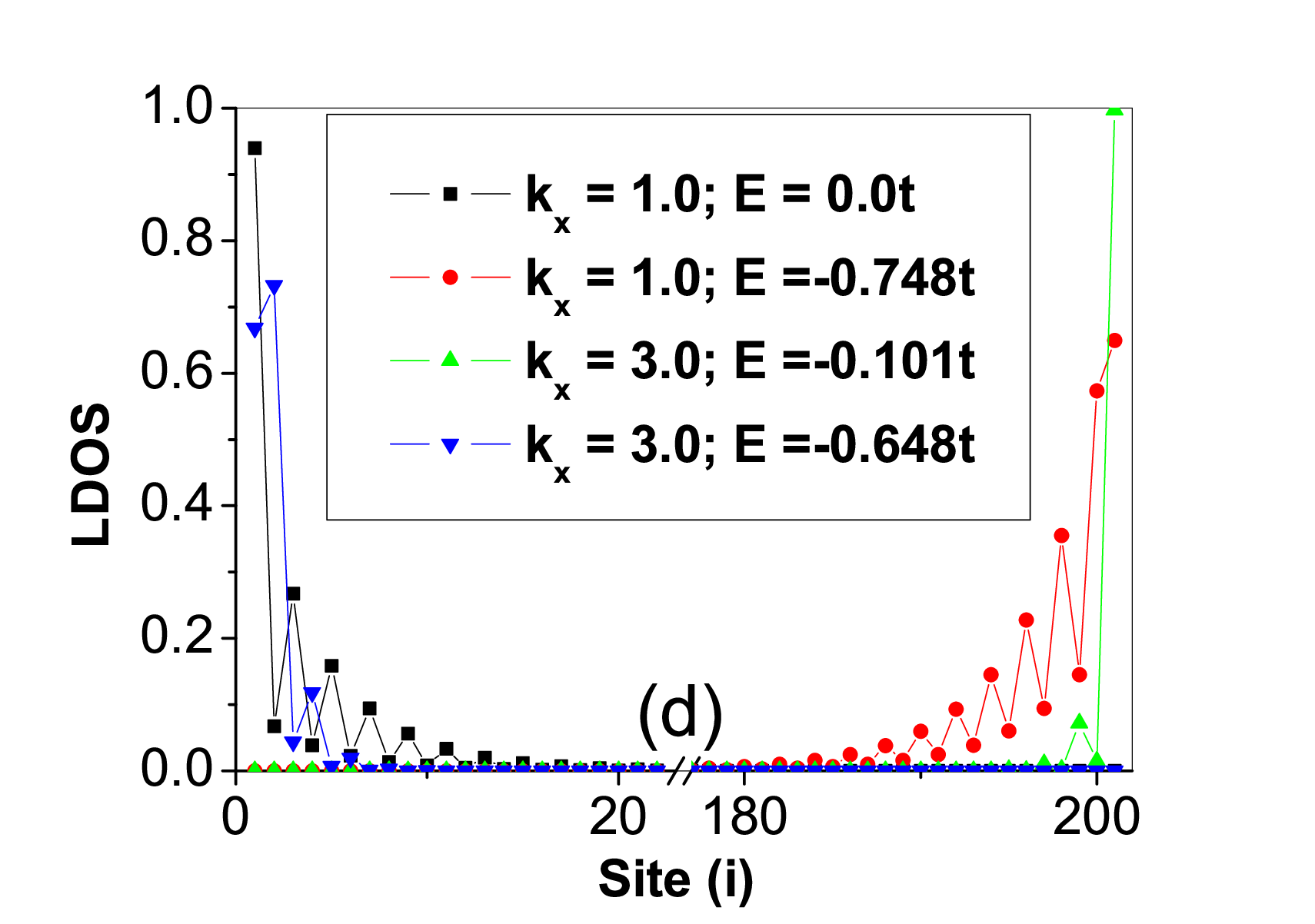}
\caption{(Color online) Energy band structures of (a)
zigzag-zigzag-, (b) bearded-bearded-, (c) zigzag-bearded-edged MGNRs, and (d)
the probability distribution $|\protect\psi|^{2}$ as functions of
position
across nanoribbon of edge states corresponding to (c). The parameter we used for (a)-(c) is $t_{0}=0.5$.} \label{fig:three}
\end{figure}

In our numerical calculations, a periodic boundary
condition (see Fig. \ref{fig:one}) is considered along the direction of the
edge of graphene nanoribbons such that the wave vector $k_{x}$($k_{y}$)
($k_{x}$ for zigzag-edged nanoribbons and $k_{y}$ for armchair-edged
nanoribbons, the corresponding length unit is $\sqrt{3}a_0$ and $3a_0$, respectively. Where $a_0\approx 0.142$nm is the carbon-carbon bonding length) along the edge direction is a good quantum number. After doing
Fourier transformation in this direction, a Hamiltonian can be
diagonalized exactly. The eigenenergies $E(k_{x})$ [$E(k_{y})$] and
eigenstates $\psi (E,k_{x},y)$ [$\psi (E,k_{y},x)$] can be calculated
numerically, from which we obtain the energy bands and probability
distribution $|\psi |^{2}$ as functions of the position across the graphene
nanoribbon for a fixed eigenenergy $E$ and a good quantum number $k_{x}$($%
k_{y}$). Results are given in Fig. \ref{fig:two} for armchair-edged MGNRs and in Fig. \ref{fig:three} for zigzag-zigzag-,
bearded-bearded-, and zigzag-bearded-edged MGNRs. Many of
these results are discussed in several other
literatures\cite{Jaroslaw,Niu,Akhmerov,Kenichi,Nakada,MFujita}. We
have presented them here to facilitate the discussion of the results
obtained for edge-hopping modulation.

The electron spectrum of MGNRs with zigzag and armchair edges, but
without edge-hopping modulation, have been extensively studied by
several methods, such as matching the boundary condition
by combining all the possible wave functions of the
Schr\"{o}dinger equation near Dirac points\cite{Akhmerov} or
discrete TBA\cite{Kenichi} and the transfer matrix
approach\cite{Nakada}. In both the zigzag- and bearded-edged MGNRs, the
zero mode of edge states appears and its wave function vanishes at
one sublattice and localizes at another sublattice with
exponential decay away from the edges. No edge state is found for
armchair-edged MGNRs, even after n.n.n. hopping is taken into
account. The zero mode of an edge state can exist in the following regions:
$2\pi /3<k_{x}<4\pi /3$ for zigzag-zigzag-edged
MGNRs\cite{MFujita}, but $-2\pi /3<k_{x}<2\pi /3$ for
bearded-bearded-edged MGNRs. If the MGNRs are zigzag-bearded-edged
MGNRs\cite{Niu}, their edge
states will spread to the entire region of $k_{x}\in (-\pi ,\pi )$. All
these results are repeated in comparison with the
ones in the presence of edge-hopping
modulation. Our calculations show that some extra edge states
emerge owning to edge-hopping modulation (see Figs. \ref{fig:two}
and \ref{fig:three}). These results are something analogous to
the edge states reported in ref.~\citen{Jaroslaw}.
In Fig.
\ref{fig:two}, we give the energy spectrum of armchair-edged MGNRs
with $t_{0}\neq t$, where (a), (b), and (c) correspond to
$t_{0}=0.2,0.5$, and $0.8$, respectively. From Figs. \ref{fig:two}(a)-\ref{fig:two}(c),
we can see that the linear dispersion near the Dirac point
$K(K^{\prime })$ is preserved, but a small gap may appear owning to
the finite width of the nanoribbons. Tuning the ratio of $t_{0}/t$
from $0.2$ to $1.0$, the edge states are changed dramatically. In
the limit of $t_{0}\rightarrow t$, the edge states completely
disappear as we expect. This clearly demonstrates that the edge
states can be manipulated by edge-hopping modulation.

Very similar pictures are also obtained for zigzag-zigzag-,
bearded-bearded- and zigzag-bearded-edged MGNRs, where some
additional edge states emerge owning to edge-hopping modulation
(see Fig. \ref{fig:three}). The probability distributions $|\psi
|^{2}$ of edge states as functions of the position across the
transversal direction of nanoribbons are presented in Figs.
\ref{fig:two}(d) and \ref{fig:three}(d). They show
exponential decay away from the edge. Figure \ref{fig:three}(a)
shows the energy spectrum of the edge states of zigzag-zigzag-edged MGNRs with edge-hopping modulation. Some extra edge states
can be found in the region of $-2\pi /3<k_{x}<2\pi /3$ in Fig. \ref{fig:three}%
(b) for bearded-bearded-edged MGNRs, and in Fig. \ref{fig:three}(c)
for
zigzag-bearded-edged MGNRs. They have similar features as shown in Fig. \ref{fig:three}%
(a). All extra edge states of MGNRs disappear in the limit of $%
t_{0}\rightarrow t$, as expected.

In fact, within the TBA, most of the lattice models
for studying edge states can be mapped to a 1D periodic lattices with
several different types of \textquotedblleft atoms\textquotedblright\ in a
unit cell. Taking zigzag-edged MG as an example, the unit cell mapped from
the zigzag-edged MG model $k_{x}(k_{y})$ to 1D lattices for a good
quantum number contains two types of \textquotedblleft
atoms\textquotedblright\ in the n.n. hopping approximation. Based on the
transfer matrix method or TBA, the problem of finding a solution of
edge states can be mapped into a $2\times 2$ matrix equation\cite%
{zhaoyuanyuan,zhaoyuanyuan2}. From the form of the wave function
of edge states: $\psi _{n}=Ce^{-\lambda na}+De^{\lambda na}$, where
$a$ is a periodic constant and $\lambda >0$. The existence of
edge states must satisfy two sufficient and necessary conditions
with $C\neq 0$ and $D=0$. The coefficients $C$
and $D$ are functions of the bulk hopping parameter $t$, the excitation energy $%
E$ of an edge state, and the wave vector $k_{x}$ (a good quantum
number). Also, they depend on the edge boundary condition. In the
case of zigzag-edged graphene without edge-hopping modulation
($t_{0}=t),$ the condition $C\neq 0$ can be satisfied in the
region of $2\pi /3<k_{x}<4\pi /3$, but $C=0$ in the region of $-2\pi
/3<k_{x}<2\pi /3$. Then the condition $D=0$ yields the energy
dispersion relation of edge states, {\it i.e.}, $E(k_{x}):k_{x}\in (2\pi
/3,4\pi /3)$. However, edge-hopping modulation can directly
change the edge boundary condition that easily breaks the equation
$C=0:$ $k_{x}\in (-2\pi /3,2\pi /3) $ and leads to the condition
$C\neq 0$; therefore the extra states can extend to all Brillouin
zones. Meanwhile, edge-hopping modulation also modifies the
coefficient $D$ that leads to the change in the energy dispersion of
edge states. Here, graphene is just an example, and the method of
edge-hopping modulation for manipulating edge states can be
applied to other systems. Thus, it might be recommendable to
manipulate the edge states for future applications.

%%%%%%%%%%%%%%%%%%%%%%%%%%%%%%%%%%%%%%%%%%%%%%%%%%%%%%%%%%%%%%%%%%%%%%%%
\section{Edge States of MGNRs with SOC}
\label{SLGNR_SOC}
%%%%%%%%%%%%%%%%%%%%%%%%%%%%%%%%%%%%%%%%%%%%%%%%%%%%%%%%%%%%%%%%%%%%%%%%

In order to clarify the effect of edge-hopping modulation on
edge states when the SOC is present, we turn to the study of Kane and Mele's model\cite{Kane1, Kane2} with the SOC system. The Hamiltonian of this model reads:

\begin{eqnarray}
{\hat{\mathcal{H}}}_{2}&=&{\hat{\mathcal{H}}}_{1}+i\frac{2\lambda _{SO}}{\sqrt{3}}
\sum_{\langle\langle i,j\rangle\rangle}\nu_{ij}c_{i}^{\dag}\hat{\sigma}\cdot(
\vec{d}_{kj}\times \vec{d}_{ik})c_{j}
\qquad\qquad\nonumber \\
&+&i\lambda_{R}\sum_{\langle i,j\rangle}c_{i}^{\dag}\vec{z}\cdot(\hat{\sigma}\times
\vec{d}_{ij})c_{j},
\label{eq:two}
\end{eqnarray}
where $\hat{\mathcal{H}}_{1}$ is that in eq. (\ref{eq:two}) the same as that in eq. (\ref{eq:one}), $%
c_{i}^{\dagger }=(c_{i\uparrow }^{\dag },c_{i\downarrow }^{\dag }),\hat{\sigma} $
the Pauli matrix, and $\vec{z}$ the unit vector along the perpendicular direction to
the plane of graphene nanoribbons. The second term is an intrinsic SOC and $\langle\langle i,j\rangle\rangle$ in the first summation represents two n.n.n. sites $%
\{i,j\},$ and the subscript $k$ denotes a unique common
n.n. site of sites $i$ and $j$. $\vec{d}_{ik}$ is the unit vector
that points from $k$ to $i$. The third term on the left side of eq. (\ref{eq:two}) is
Rashba SOC where the summation is carried over all n.n. sites $\langle i,j\rangle$.
The constant $\lambda _{SO}$ can be estimated using the perturbation theory or
first-principles calculation. Its range is from $0.001$meV to $0.05$meV. The
Rashba coupling $\lambda_{R}$ can be experimentally determined by
spin angle-resolved photoemission spectroscopy\cite{AVarykhalov} or
spin relaxation measurement\cite{NTombros,NTombrosO}.

\begin{figure}[tbp]
\centering
\includegraphics[bb=1 1 540 430,width=4.1cm,height=3.4cm]{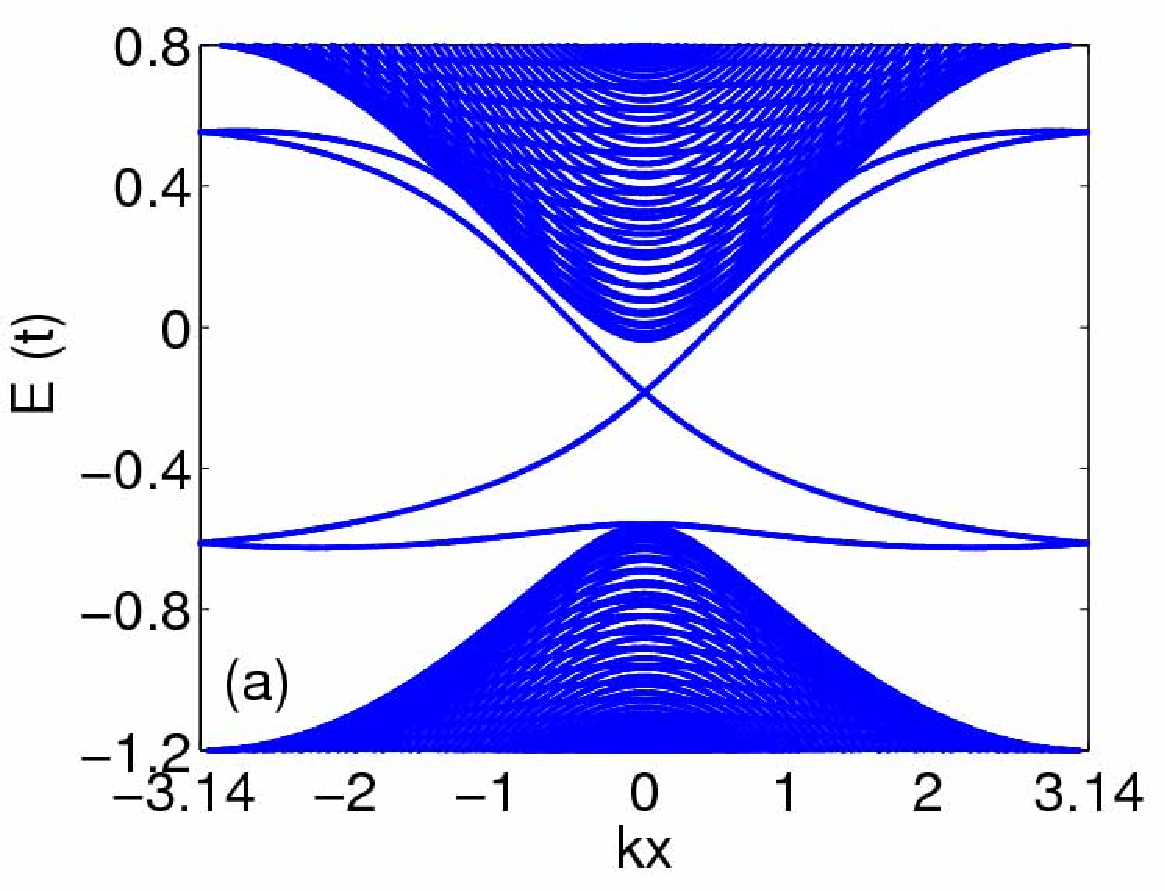} \hspace{1mm}
\includegraphics[bb=1 1 540 430,width=4.1cm,height=3.4cm]{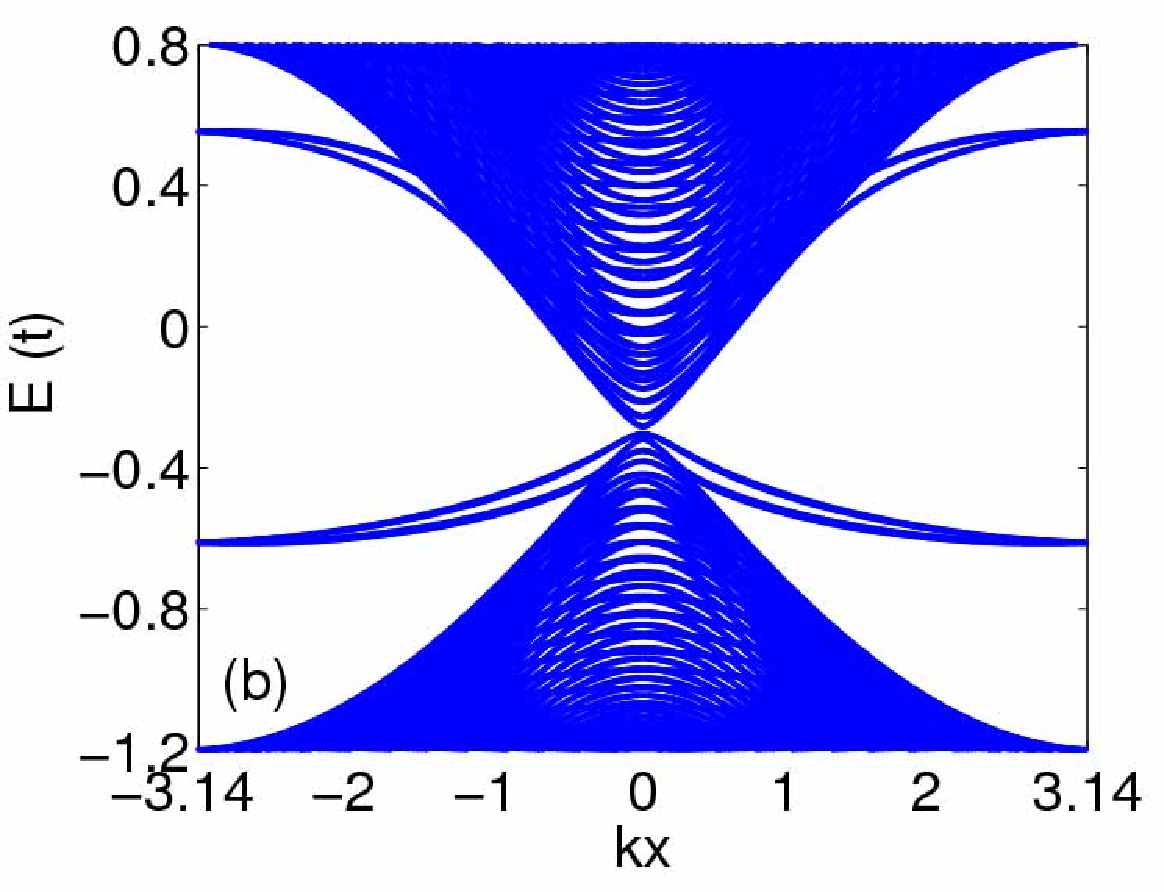} \hspace{1mm} %
\includegraphics[bb=1 1 540 430,width=4.1cm,height=3.4cm]{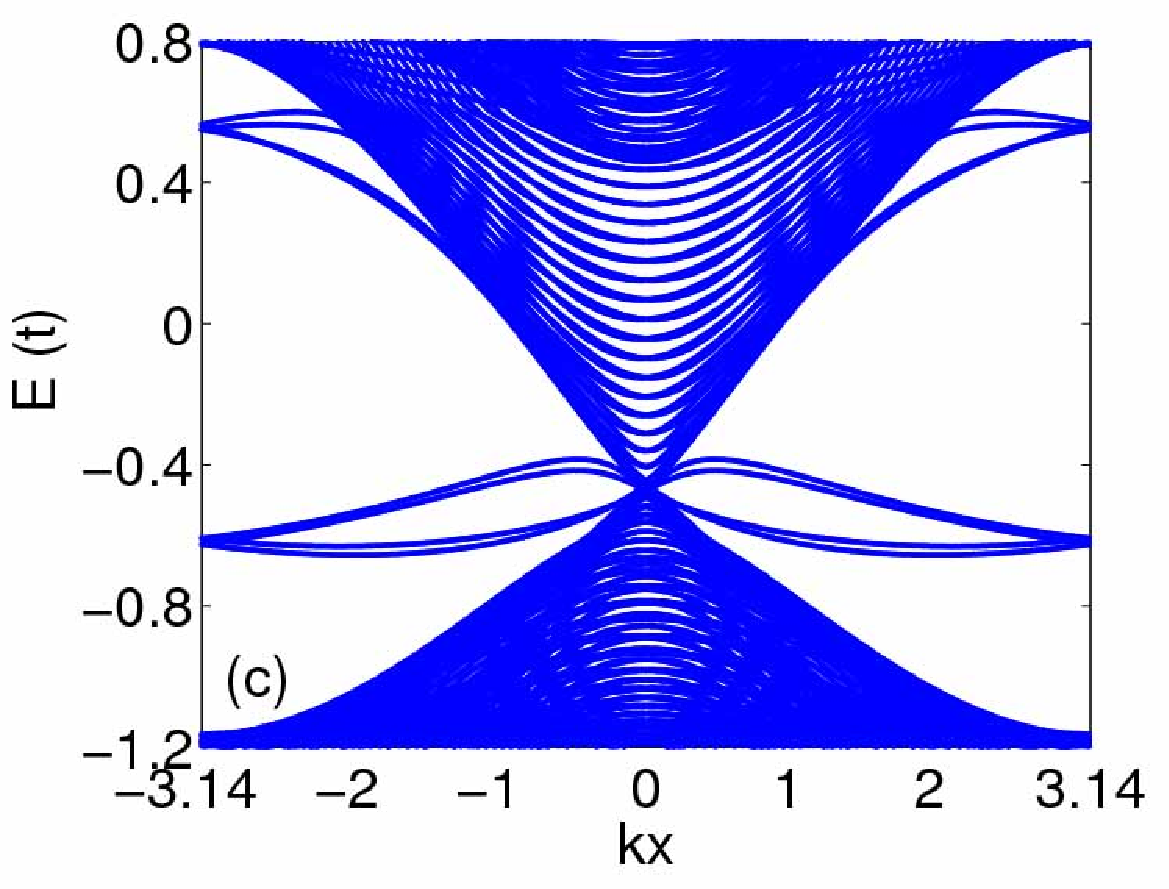}
\hspace{1mm}
\includegraphics[bb=9 1 730 534,width=4.0cm,height=3.4cm]{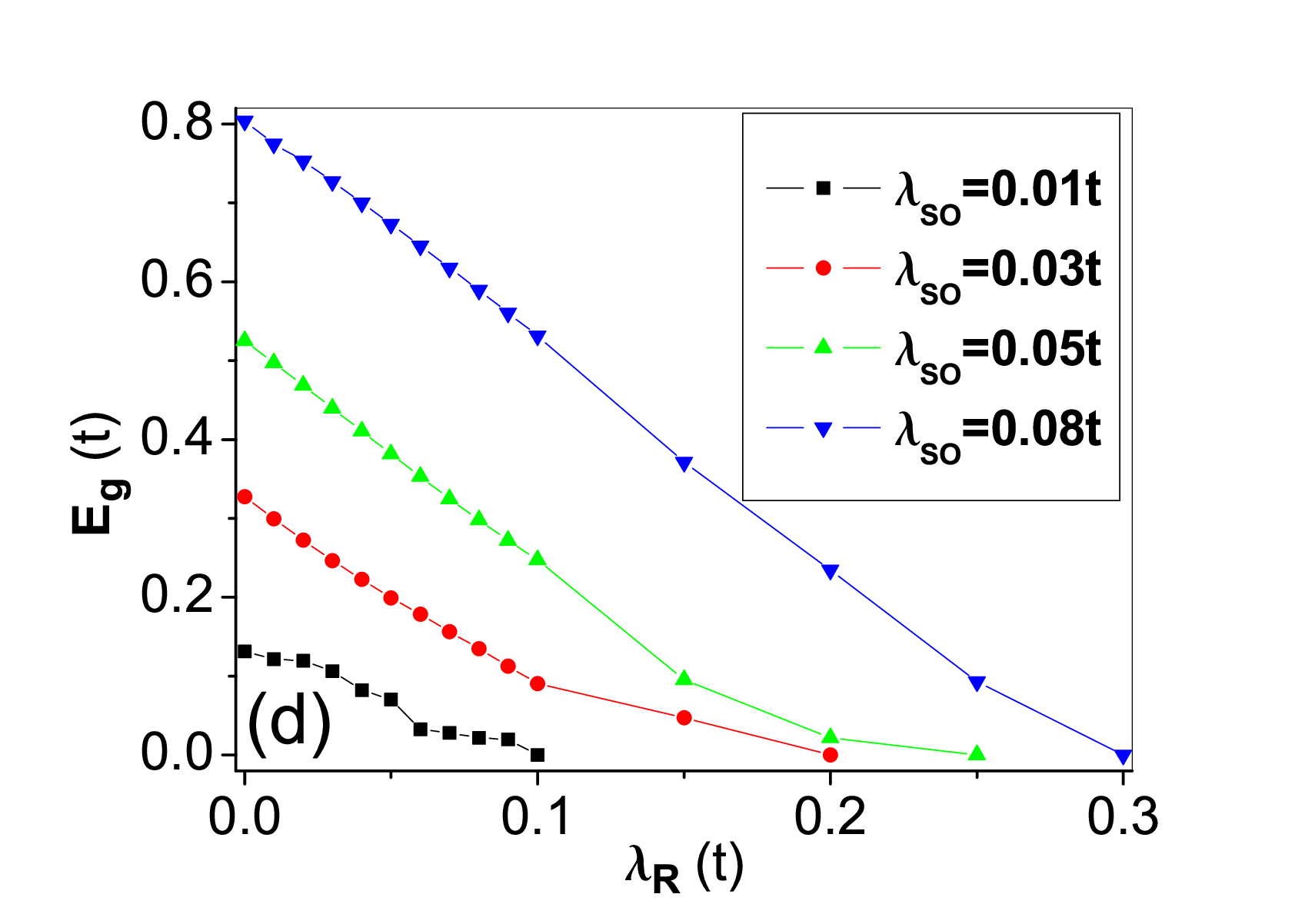}
\caption{(Color online) (a)-(c) Energy band structure of the armchair-edged
MGNRs, and (d) energy gap $E_{g}$ as a function of $\protect\lambda_{R}$ at different values of $\protect\lambda_{SO}$. The parameters we used are as follows:
for (a), $\protect\lambda_{SO} = 0.05$ and $\protect\lambda_{R} =
0$; for (b), $\protect\lambda_{SO} = 0$ and $\protect\lambda _{R} = 0.05$; for (c), $\protect\lambda_{SO} = 0.05$ and $\protect\lambda_{R} =
0.2$.} \label{fig:four}
\end{figure}

The effects of Rashba and intrinsic SOC in graphene nanoribbons
with edge-hopping modulation are presented in Fig. \ref{fig:four}. We find some
extra edge states passing through the bulk energy gap, where spin
degeneracy has been split, except at the crossing point on which degeneracy is
protected by time reversal symmetry (TRS: Kramers degeneracy) even in the presence of edge-hopping modulation. The bulk is
an insulator, whereas the edge is a metal irrespective of the
geometric structure of the edge, {\it i.e.}, armchair, zigzag, or bearded. It is shown that edge states inside an
energy gap are robust against edge-hopping modulation in the presence
of SOC. Some new extra edge states are created outside the
energy gap by edge-hopping modulation, and the energy dispersion relation
is slightly changed. If only the Rashba SOC $\lambda _{R}$ exists, there is no bulk energy gap
and it is not an insulator in the bulk. Differently from Rashba SOC,
intrinsic SOC is favorable for opening a bulk energy gap around Dirac points\cite{Ralph}.

Therefore, there are two conclusions that can be drawn:

$1).$ Rashba SOC depresses the gap and it is unfavorable for the
topological phase of QSHE. Quantitatively, the energy band gap
$E_{g}$ as a function of the Rashba SOC $\lambda _{R}$  has an
approximately linear relation, {\it i.e.}, $E_{g}\sim(\lambda
_{R}^{c}-\lambda _{R})$, which is shown in Fig.
\ref{fig:four}(d). There is a critical value of $\lambda _{R}^{c}$
at which the band energy gap induced by intrinsic SOC is closed when
$\lambda _{R}\geq \lambda _{R}^{c}$. The larger the $\lambda _{SO}$,
the higher the critical value of $\lambda _{R}^{c}$ will be.

$2).$ From Figs. \ref{fig:four}(a)-\ref{fig:four}(c), edge states always
exist owning to SOC irrespective of the existence of the bulk
energy gap. When the spin Chern number\cite{DNSheng} reaches
to zero, it shows that the bulk energy is closed, but that edge
states also exist.

%%%%%%%%%%%%%%%%%%%%%%%%%%%%%%%%%%%%%%%%%%%%%%%%%%%%%%%%%%%%%%%%%%%%%%%%
\section{Edge States in BGNRs with SOC}
\label{BLGNR_SOC}
%%%%%%%%%%%%%%%%%%%%%%%%%%%%%%%%%%%%%%%%%%%%%%%%%%%%%%%%%%%%%%%%%%%%%%%%

Now let us discuss BGNRs with and without SOC in this section. Owning to the fact that edge states are robust against edge-hopping modulation at low-energy excitations, we neglect the effect of edge-hopping modulation here. Intrinsic SOC in MG favors the opening of an energy
gap\cite{Ralph}, which is very small. However, a significant
energy gap can be opened and tuned in BG using an external bias
field\cite{Eduardo,CASTRO}. In this section, we will study what
happens for bias voltage BGNRs in the presence of SOC.
Although the true intrinsic SOC effect in graphene is small, we
would like to address the problem in an artificial model, with
the expectation that the theoretical results in this section will be
helpful in understanding the physics of few-layer hexagonal
systems.

The geometric structure of BGNRs is expressed in Fig. \ref{fig:one}(b). The
Hamiltonian of BGNRs with SOC is studied as follows:

\begin{figure}[tbp]
\centering
\includegraphics[bb=1 1 540 430,width=4.1cm,height=3.4cm]{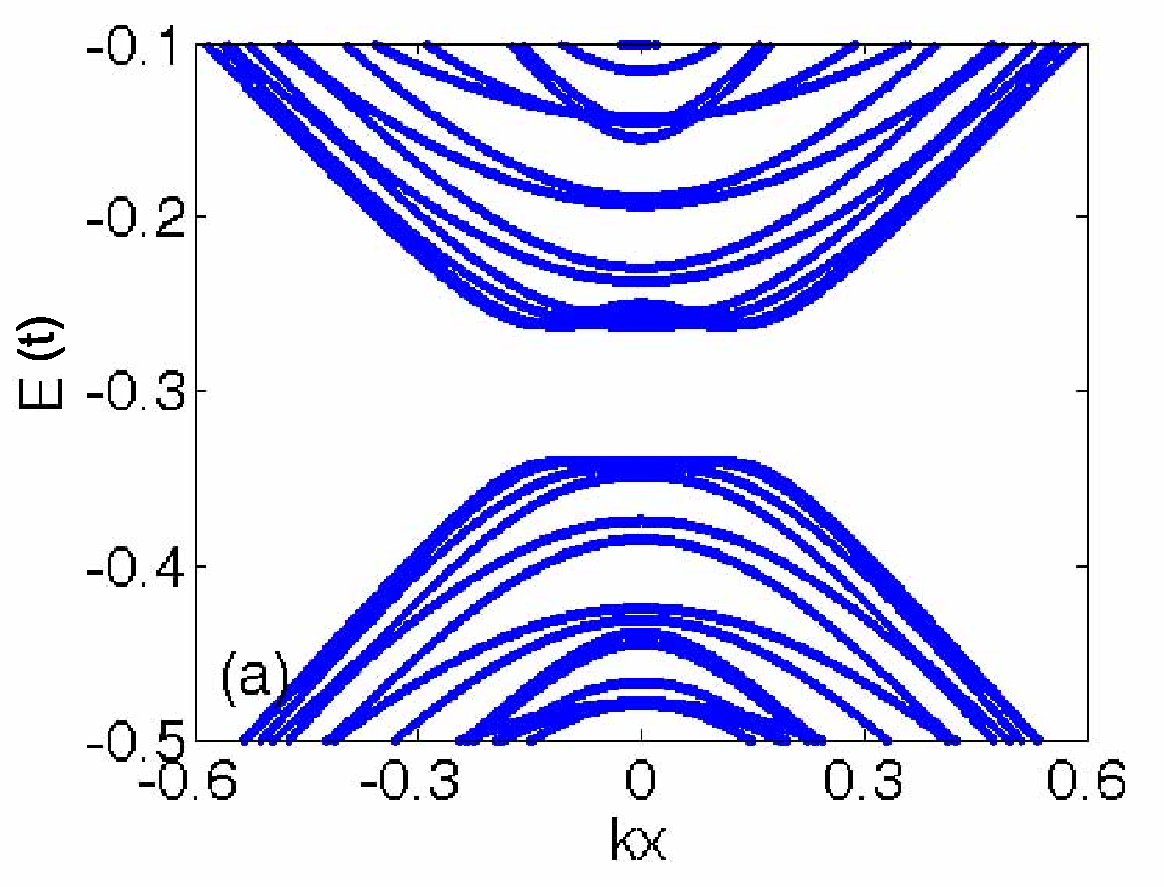} \hspace{1mm}
\includegraphics[bb=1 1 540 430,width=4.1cm,height=3.4cm]{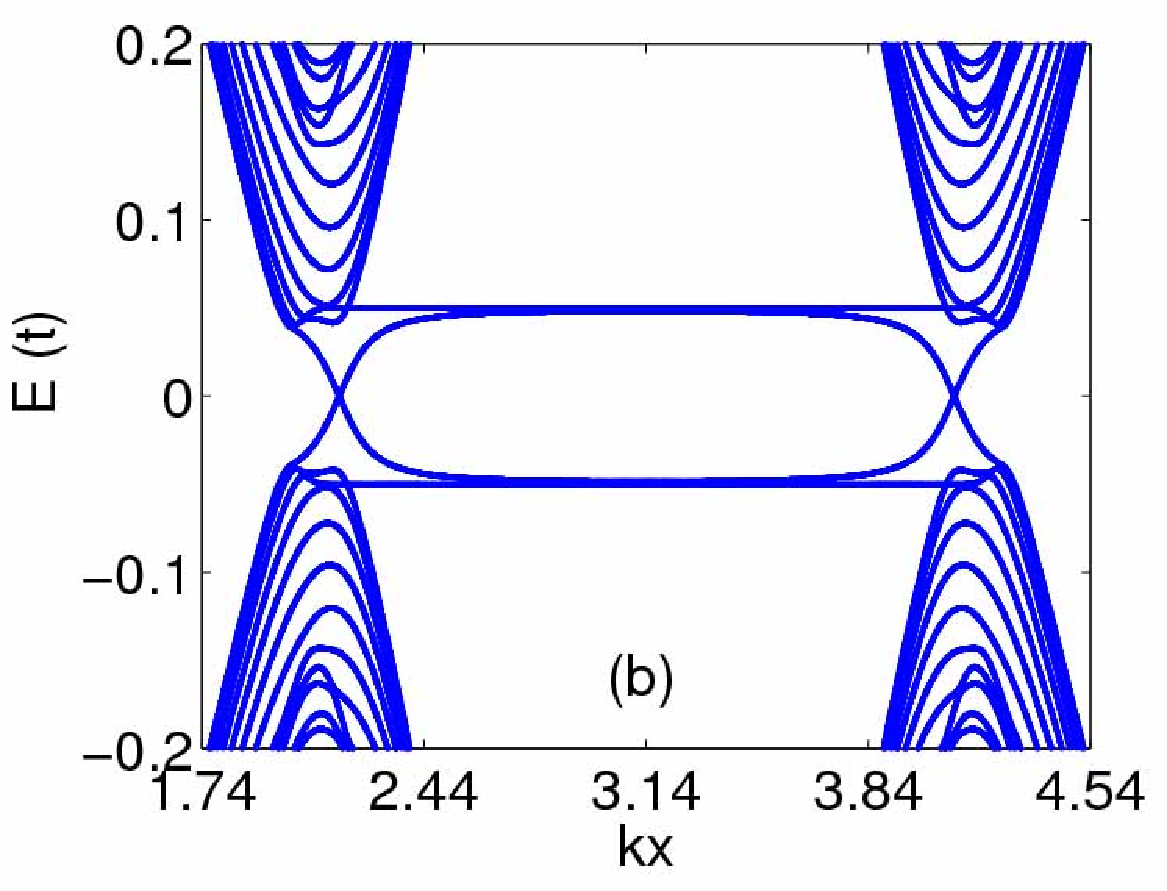} \hspace{1mm} %
\includegraphics[bb=1 1 540 430,width=4.1cm,height=3.4cm]{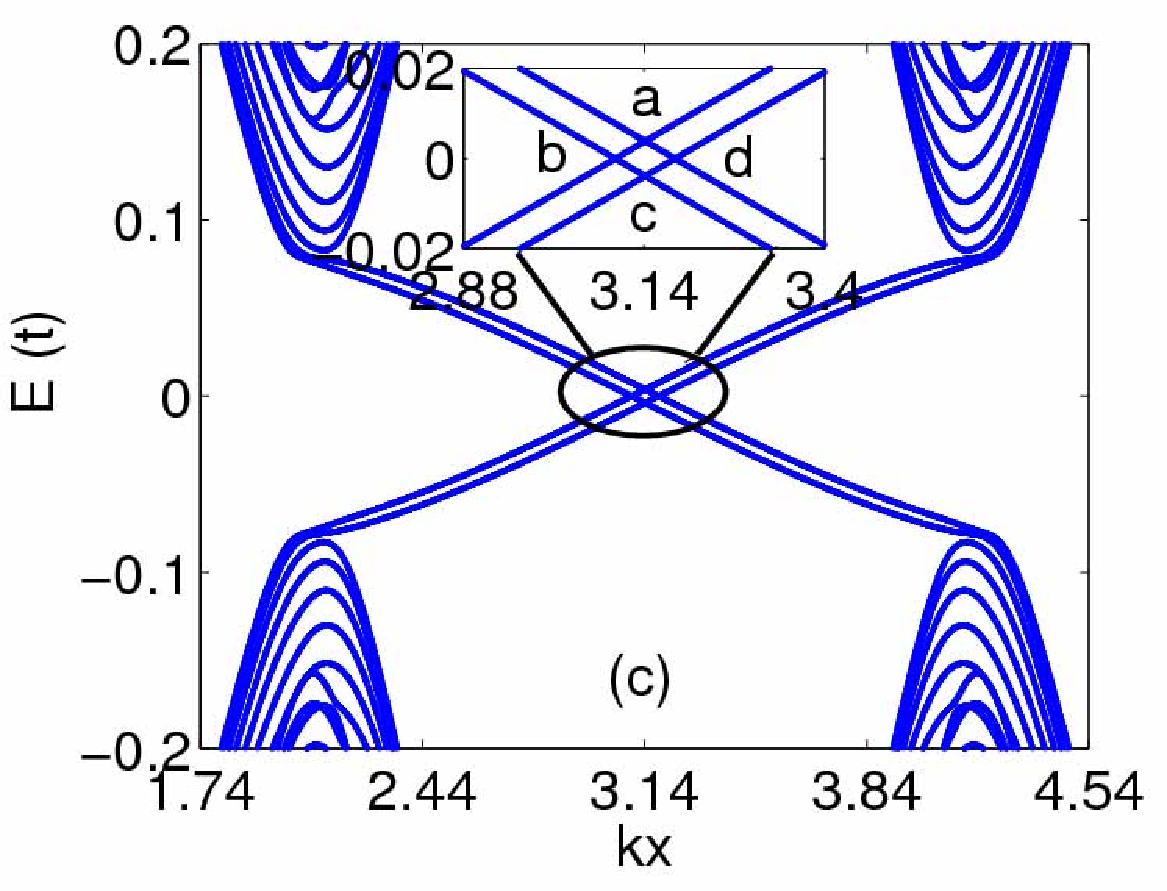}
\hspace{1mm}
\includegraphics[bb=1 1 540 430,width=4.1cm,height=3.4cm]{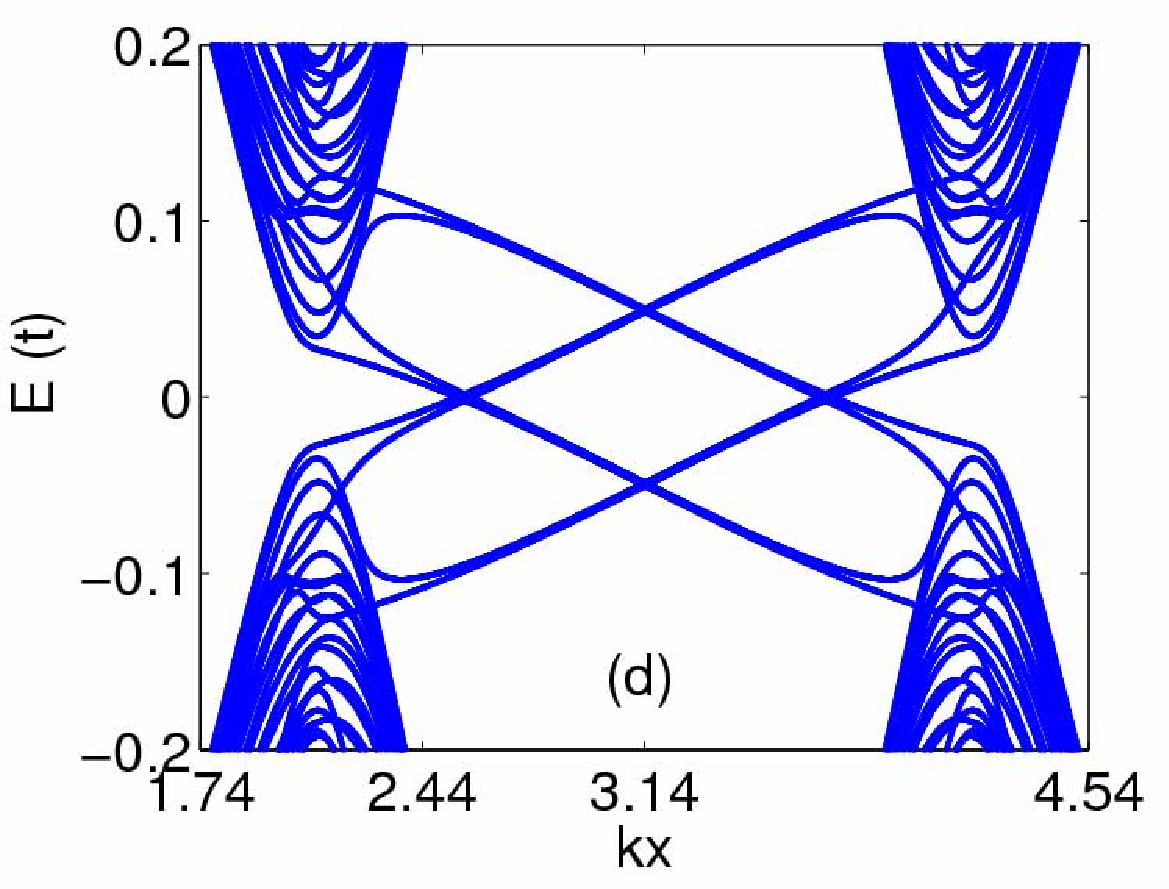}
\caption{(Color online) Energy band structure of BGNRs. In (a), it is the armchair-edged system with n.n.n. intra- and inter-layer
hoppings, and in (b)-(d) it is the zigzag-edged system without n.n.n. intra- and inter-layer
hoppings. The inset in (c) is a magnified bulk gap region near $k_x=\pi$. The parameters we used are as follows: for (a)-(b), $\protect\lambda_{SO} = 0$ and $V_{B} = 0.1$; for (c), $\protect\lambda_{SO} = 0.015$ and $V_{B} = 0.0$; and
for (d), $\protect\lambda_{SO} = 0.015$ and $V_{B} = 0.1$.}
\label{fig:five}
\end{figure}

\begin{eqnarray}
{\hat{\mathcal{H}}} &=&\sum_{\delta }{\hat{\mathcal{H}}}_{2\delta }+i\lambda
_{R}^{\perp }\sum_{\ll i,j\gg ,\alpha \beta }c_{i,\alpha ,1}^{\dag }({\hat{s}%
}\times {\hat{d}}_{ij})_{z}c_{j,\beta ,2}
\qquad\qquad\nonumber \\
&+&t_{\bot }\sum_{<i,j>,\sigma
}c_{i,\sigma ,1}^{\dag }c_{j,\sigma ,2}
+t_{\bot }^{\prime }\sum_{\ll i,j\gg ,\sigma }c_{i,\sigma ,1}^{\dag
}c_{j,\sigma ,2}
\qquad\qquad\nonumber \\
&+&\frac{V_{B}}{2}\sum_{i,{\hat{\sigma}}}\left( c_{i,\sigma {,1%
}}^{\dag }c_{j,\sigma {,1}}-c_{i,\sigma {,2}}^{\dag }c_{j,\sigma {,2}%
}\right),\label{eq:three}
\end{eqnarray}%
where $\hat{\mathcal{H}}_{2\delta }$ is the Hamiltonian of two
MGs without interlayer coupling and each MG is described by eq. (\ref%
{eq:two}) with an additional layer index $\delta (=1,2)$ to
distinguish the creation/annihilation operators at different
layers. Both intrinsic and intralayer Rashba SOCs are the same
as described in the last section. The second term of the above
Hamiltonian describes the n.n.n. interlayer Rashba SOC induced by a
bias voltage $V_{B}$ across two layers of BGNRs since the effective
potential gradient crossing two layers of BGNRs may be largely
contributed by the bias voltage. The third and fourth terms give the
n.n. and n.n.n. hoppings of interlayer\cite{DNShengBilayer} sites
in typical Bernal stacking, respectively. The last term
describes the potential difference due to the bias voltage $V_{B}$.

\begin{figure}[tbp]
\centering
\includegraphics[bb=1 1 540 430,width=4.1cm,height=3.4cm]{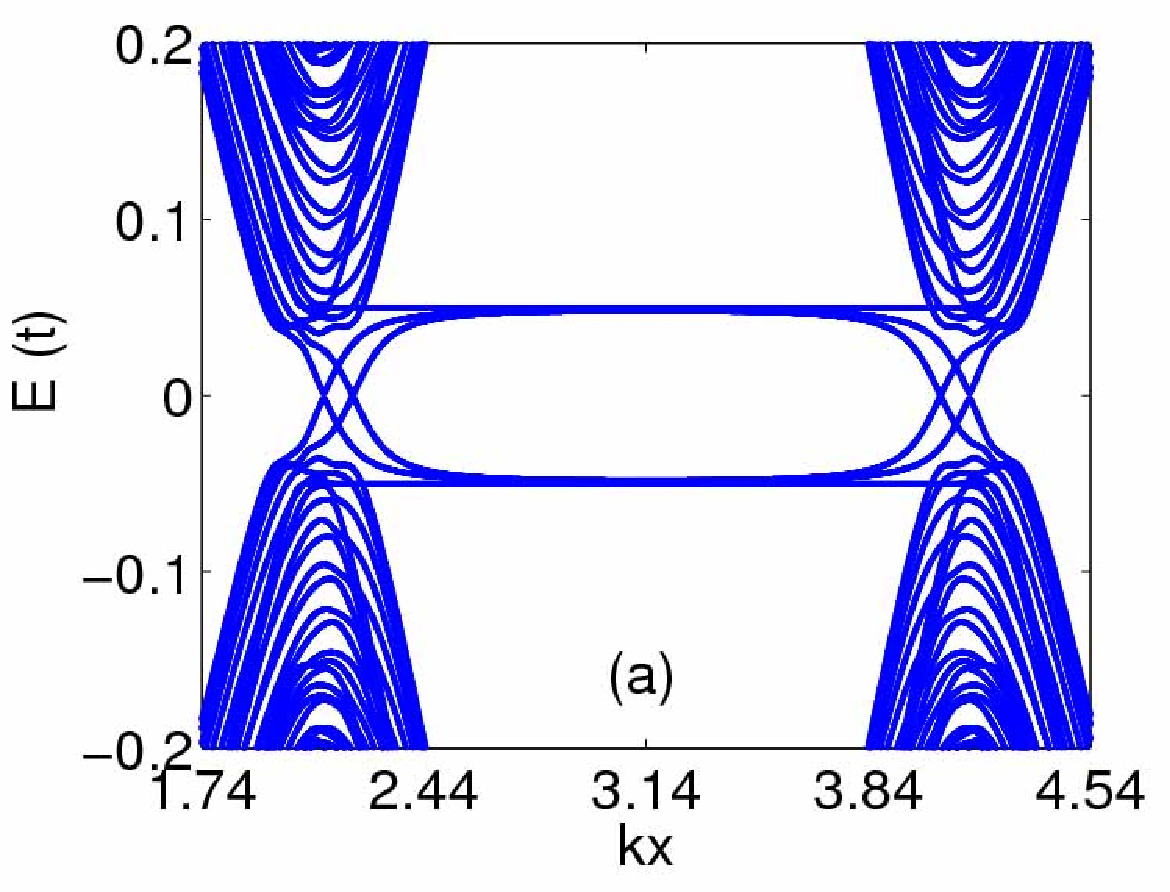} \hspace{1mm}
\includegraphics[bb=1 1 540 430,width=4.1cm,height=3.4cm]{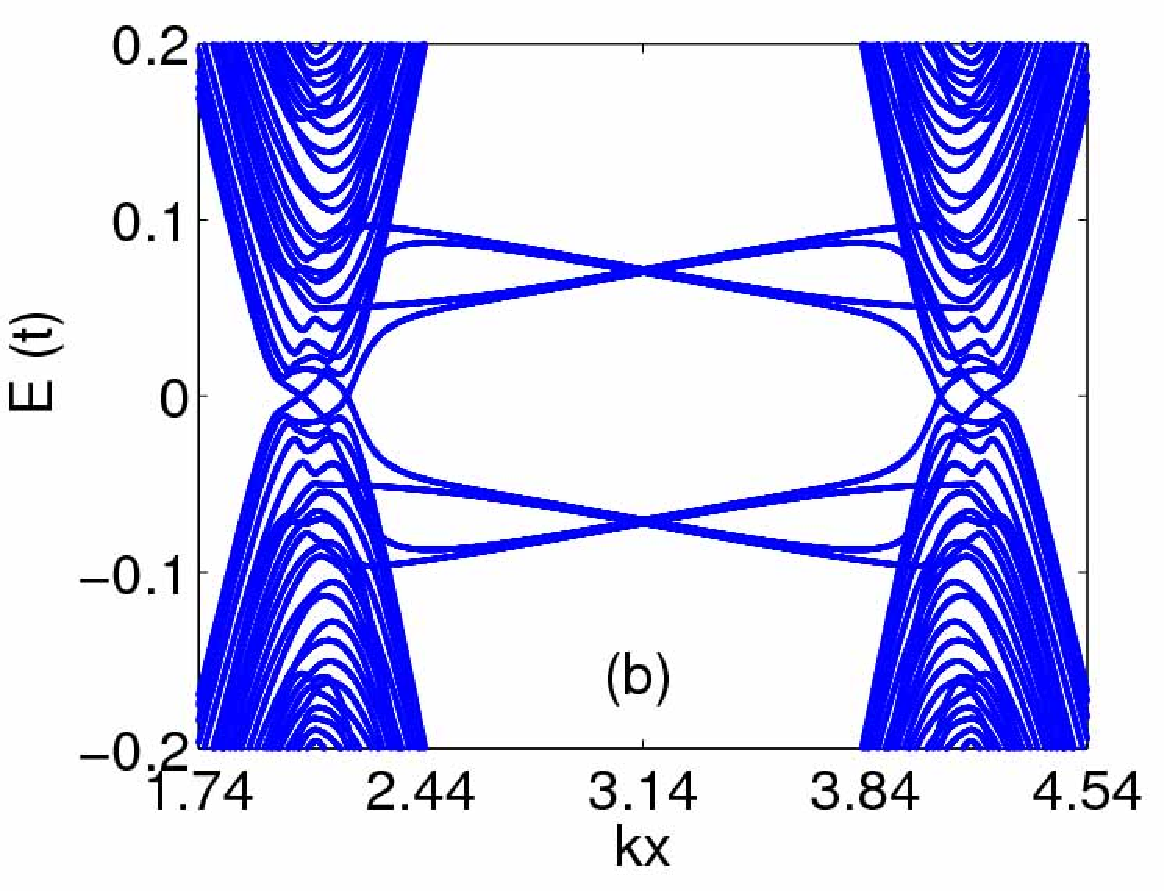} \hspace{1mm} %
\includegraphics[bb=1 1 540 430,width=4.1cm,height=3.4cm]{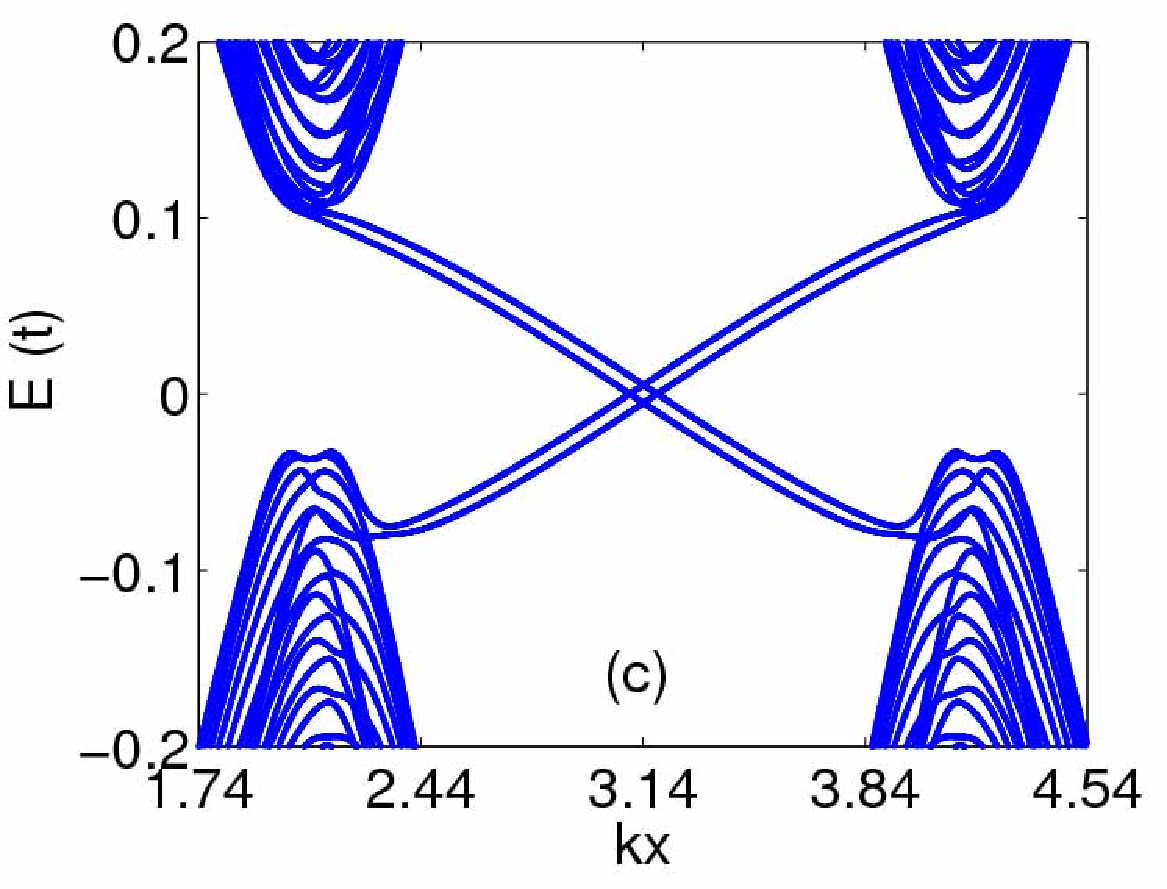}
\hspace{1mm}
\includegraphics[bb=1 1 540 430,width=4.1cm,height=3.4cm]{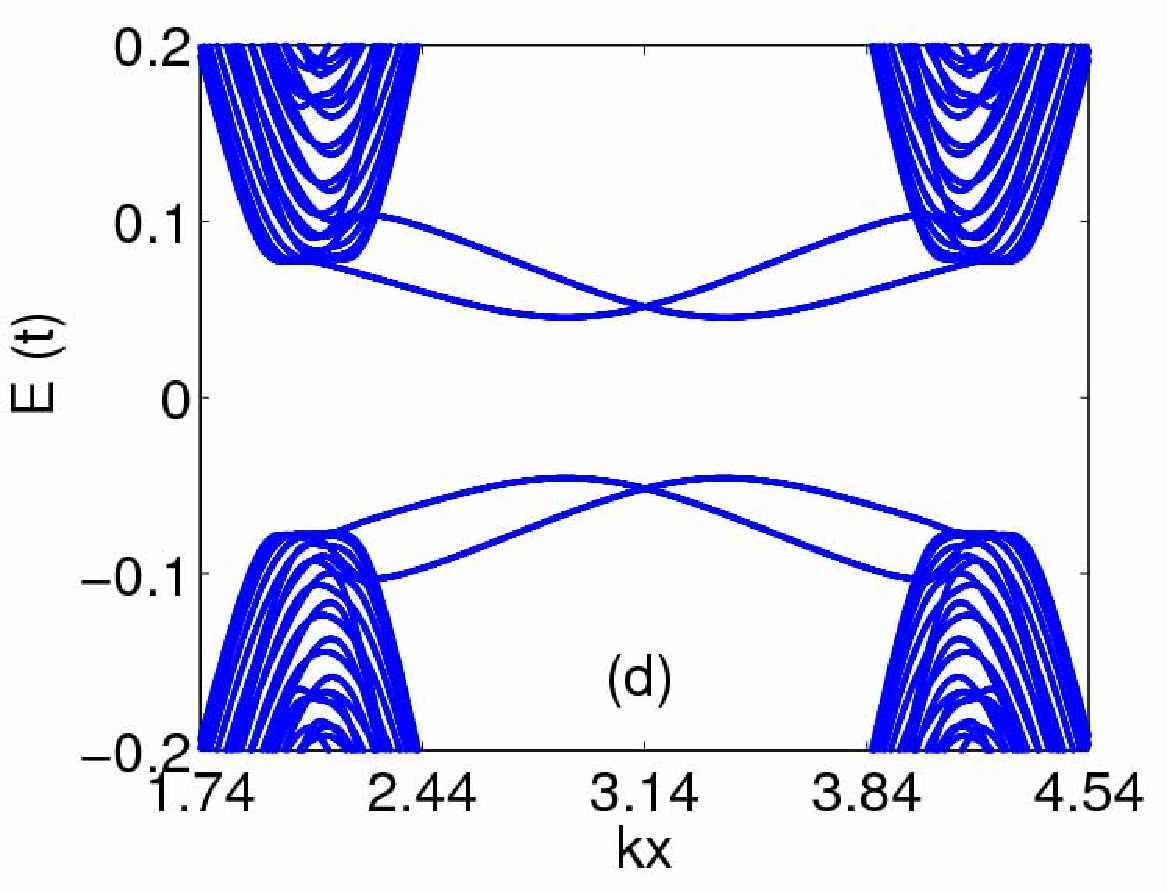}
\caption{(Color online) Energy band structure of the zigzag-edged BGNRs. In (a) and (b), the system has bias voltage but without intrinsic SOC,
and in (c) and (d) the system has intrinsic SOC but without bias voltage. The parameters we used are as follows: for (a) and (c), $\protect\lambda_{R} = 0.03$ and $\protect\lambda^{\perp}_{R} = 0$; and for (b) and (d), $\protect\lambda_{R} = 0$ and $\protect\lambda^{\perp}_{R} = 0.03$.}
\label{fig:six}
\end{figure}

The energy spectrums of BGNRs are given in Fig. \ref{fig:five}(a)
for armchair-edged BGNRs and Figs. \ref{fig:five}(b)-\ref{fig:five}(d) for zigzag-edged BGNRs. It is shown that not any edge state can be found for
armchair-edged BGNRs [see Fig. \ref{fig:five}(a)]. However, there are
edge states of zigzag-edged BGNRs [see Fig. \ref{fig:five}(b)]
where two degenerate edge modes in decoupled zigzag-edged BGNRs
have been split and mixed owning to the interlayer coupling
\cite{CastroPRL}. The n.n.n. hopping included is just for breaking the
electron-hole symmetry and leads to a change in the energy dispersion.
The linear dispersive spectrum of BGNRs
around the Dirac point changes to a parabolic one owning to
interlayer coupling and no energy gap appears in the band if no bias
voltage is applied. A bias voltage leads to the opening of a band gap [see
Figs. \ref{fig:five}(a) and \ref{fig:five}(b)]. In Figs. \ref{fig:five}(c) and
\ref{fig:five}(d), the band energy gap is opened by intrinsic SOC, and edge
states exist for zigzag-edged BGNRs. The difference between
Figs. \ref{fig:five}(b) and \ref{fig:five}(c) is that the band
gap in Fig. \ref{fig:five}(b) is induced by the bias voltage $V_{B}$,
but that in Fig. \ref{fig:five}(c) by intrinsic SOC. Figure \ref%
{fig:five}(c) clearly shows that two Dirac cones displace the up-down shift
due to interlayer coupling. The TRS at the crossing points $a$ and $c$ in
Fig. \ref{fig:five}(c) is preserved, but the degeneracy at the two crossing
points $b$ and $d$ is accidental and not conserved by TRS. Thus, it can be
easily split by the perturbation of the interlayer Rashba SOC $\lambda
_{R}^{\bot }$ shown in Fig. \ref{fig:six}(d).

From Figs. \ref{fig:five}(c) and \ref{fig:five}(d), we can see that the
band energy gap opened by intrinsic SOC will be reduced gradually for zigzag-edged BGNRs if the bias voltage increases. It is interesting to point out
that the bias voltage is favorable for opening the band energy gap without
intrinsic SOC. However, it will become a negative factor to keep a band
gap in the presence of intrinsic SOC.

In Fig. \ref{fig:six}, we discuss the competition among Rashba SOC,
intrinsic SOC, and bias voltage. It is shown that both intralayer and
interlayer Rashba SOCs induce spin polarization, but decrease the
band energy gap [see Figs. \ref{fig:six}(a) and \ref{fig:six}(b)]. The band energy gap is
opened by the bias voltage $V_{B}$. The difference between Figs. \ref{fig:six}(c)
and \ref{fig:six}(d) shows that the interlayer Rashba SOC $\lambda _{R}^{\bot }$ makes
the edge state band gapful. However, the Dirac points $a$ and $c$ in Fig.
6(c) are still preserved by TRS. It shows that two individual topological
MG layers with $Z_{2}$ symmetry becomes a band insulator once they are coupled
together.

%%%%%%%%%%%%%%%%%%%%%%%%%%%%%%%%%%%%%%%%%%%%%%%%%%%%%%%%%%%%%%%%%%%%%%%%
\section{Conclusions}
\label{conclusions}
%%%%%%%%%%%%%%%%%%%%%%%%%%%%%%%%%%%%%%%%%%%%%%%%%%%%%%%%%%%%%%%%%%%%%%%%

In this paper, we systemically studied edge states in MGNRs and
BGNRs. We first demonstrate that some new edge states can be
created and manipulated by edge-hopping modulation for MGNRs. From this finding, it is suggested that edge-hopping modulation can be
considered as an effective way of manipulating edge states.
Then, we show that intralayer Rashba SOC  will destroy the
topological phase by suppressing the bulk energy gap opened
by intrinsic SOC for both MGNRs and BGNRs. Similarly, the bias
voltage on BGNRs can also change the energy spectrum by reducing
the energy gap in the presence of intrinsic SOC. In contrast, interlayer Rashba SOC in BGNRs can destroy the topological phase
of QSHE abruptly by opening a gap in an edge state spectrum
within a band gap, where quantum phase transition occurs and
drives the topological insulator to a band insulator.

%==================================================
\begin{acknowledgments}
This work is supported by the National Natural Science Foundation of China
under grant No. 10847001 and the National Basic Research Program of China (973
Program) under grant No. 2009CB929204 and No. 2011CB921803. W.L. acknowledges the support by the Fudan Research Program on Postgraduates and thanks Yan Chen, C. S. Ting, Qingfeng Sun, Yongjin Jiang, Jiadong Zang, and Xianlong Gao for helpful discussions. The authors are grateful to the Physical Society of Japan for financial support in publication.
\end{acknowledgments}
%==================================================

%**********************************************************************
% References
%**********************************************************************

\end{document}